\documentclass[aps,prb,superscriptaddress,twocolumn,showpacs]{revtex4}
\usepackage{amsmath,bm}
\usepackage{graphicx,epsfig,braket,amssymb,amsfonts,color}

\newcommand{\ba}{\begin{eqnarray}}
\newcommand{\ea}{\end{eqnarray}}
\newcommand{\be}{\begin{equation}}
\newcommand{\ee}{\end{equation}}
\newcommand{\bd}{\begin{displaymath}}
\newcommand{\ed}{\end{displaymath}}
\renewcommand{\v}[1]{{\bf #1}}
\newcommand{\bpm}{\begin{pmatrix}}
\newcommand{\epm}{\end{pmatrix}}
\newcommand{\nn}{\nonumber \\}

\begin{document}

\title{Three-Band Model for Quantum Hall and Spin Hall Effects}

\author{Gyungchoon Go}
\email[Electronic address:$~~$]{gcgo@skku.edu}
\affiliation{Center for Nanotubes and Nanostructured Composites, Sungkyunkwan University, Suwon 440-746, Korea}
\author{Jin-Hong Park}
\email[Electronic address:$~~$]{astatina@skku.edu}
\affiliation{Department of Physics and BK21 Physics Research
Division, Sungkyunkwan University, Suwon 440-746, Korea}
\author{Jung Hoon Han}
\email[Electronic address:$~~$]{hanjh@skku.edu}
\affiliation{Department of Physics and BK21 Physics Research
Division, Sungkyunkwan University, Suwon 440-746, Korea}
\affiliation{Asia Pacific Center for Theoretical Physics, POSTECH,
Pohang, Gyeongbuk 790-784, Korea}

\begin{abstract}
Topological properties of a certain class of spinless three-band Hamiltonians are shown
to be summed up by the Skyrmion number in momentum space, analogous to the case of
two-band Hamiltonian. Topological tight-binding Hamiltonian on a Kagome lattice is
analyzed with this view. When such a Hamiltonian is ``folded", the two bands with
opposite Chern numbers merge into a degenerate band exhibiting non-Abelian gauge
connection. Conserved pseudo-spin current operator can be constructed in this case and
used to compute the pseudo-spin Hall conductance. Our model Hamiltonians belong to the
symmetry class D and AI according to the ten-fold classification scheme.
\end{abstract}
\maketitle

\section{Introduction}

Topological properties embedded within a band structure have emerged as one of the
central themes of condensed matter physics nowadays. The topological nature is often
expressed as the gauge field derived from the underlying wave functions and manifests
itself in such observable phenomena as anomalous and spin Hall effects in
metals~\cite{nagaosa-review,niu-review}. Quantized versions of the phenomena exist in
insulators, when a certain topological number can be associated with the completely
filled bands. Classification of the permissible topological numbers in a given band
Hamiltonian was recently carried out~\cite{tenfold-way}. Numerous efforts are being made
at the moment to suggest specific models belonging to a particular entry in the
classification table with relevance to realizable experimental systems. A most striking
instance of this kind of effort is Haldane's proposal of microscopic two-band Hamiltonian
with spontaneous quantized Hall conductance~\cite{haldane}, its generalization to
time-reversal-invariant quantum spin Hall phase~\cite{kane-mele-QSH,BHZ}, and its
subsequent realization in HgTe/CdTe heterostructure~\cite{BHZ,molenkamp}.

Basically, spontaneous quantum Hall models and a subset of quantum
spin Hall models defined as two time-reversal copies of the quantum
Hall model can be phrased as a property of the generic two-band
Hamiltonian $H=\sum_{\v k} \psi^\dag_{\v k} {\cal H}_{\v k} \psi_{\v
k}$, where $\psi_{\v k}$ is a two-component spinor and the matrix
${\cal H}_{\v k}$ is ${\cal H}_{\v k} =\varepsilon_{\v k}\mathbb{I}
+ \v d_{\v k}\cdot \bm\sigma$ ($\bm \sigma$=Pauli matrix). As long
as the $\v d_{\v k}$-vector remains nonzero over the entire
Brillouin zone (BZ) such that a unique unit-vector $\hat{d}_{\v k} =
\v d_{\v k}/|\v d_{\v k} |$ exists, the two eigenstates of ${\cal
H}_{\v k}$ will be classified according to their spin helicity
$\hat{d}_{\v k}\cdot \bm \sigma | \psi^{\pm}_{\v k} \rangle = \pm|
\psi^{\pm}_{\v k} \rangle$. From the Abelian gauge connection $\v
a_{\v k} = -i \langle \psi^+_{\v k}|\bm \nabla_{\v k} |\psi^+_{\v k}
\rangle = +i \langle \psi^-_{\v k}|\bm \nabla_{\v k} |\psi^-_{\v k}
\rangle$ the associated flux density can be written in two
equivalent forms,

\ba \rho_{\v k} = \left( {\partial a_y \over
\partial k_x } - {\partial a_x \over
\partial k_y }\right) = {1\over 2} \hat{d}_{\v k} \cdot \left(
{\partial \hat{d}_{\v k} \over \partial k_x} \times {\partial \hat{d}_{\v k} \over
\partial k_y } \right). \label{eq:Skyrmion-density}\ea
Quantized number for the band, obtained as the integral $(1/2\pi)\int dk_x dk_y$ of Eq.
(\ref{eq:Skyrmion-density}), can therefore be interpreted as either the first Chern
number or the Skyrmion number of the given band~\cite{volovik,dhlee,qi} depending on the use of
the second or the third term of Eq. (\ref{eq:Skyrmion-density}) for the integrand.

There are quite a few other models embodying nontrivial topological
numbers that involve the use of 4$\times$4 $\Gamma$-matrices, many
of which are summarized in Ref.~\onlinecite{tenfold-way}. Although
the general classification scheme makes no reference to the
dimensionality of the matrix itself, in practice almost all explicit
examples of band Hamiltonians with nontrivial topology take the form
of an even-dimensional matrix in momentum space. In sharp contrast,
general discussion of the topological character for odd-dimensional
3$\times$3 matrices, or three-band Hamiltonians, appear to be
lacking. The work of Ohgushi, Murakami, and Nagaosa (OMN)~\cite{OMN}
provided, to the authors' knowledge, the first microscopic example
of a three-band model Hamiltonian with non-trivial topological Chern
number. The three-sublattice structure of the Kagome lattice is a
natural platform for the three-dimensional Hamiltonian matrix to
arise.

In this paper, we divide the general three band Hamiltonian into the
spin-1 part and nematic part then try to solve each part separately.
The spin-1 model and nematic model involve quantum Hall effect and
spin Hall effect respectively. In contrast to two band model, for
the spin-1 model, the first Chern number is one half of skyrmion
number. We realize that the existence of half-skyrmion (meron)
configuration for a unit Chern number band. When we consider the
Kagome lattice models for the nematic model, which the spin matrices
represent the pseudo-spin from the three-sublattice structure, our
spin Hall effect describes the pseudo-spin Hall effect.

The remaining part of this paper is organized as follows. In Sec.
II, in terms of spin-1 matrices, we consider a particular three band
model involving the quantum Hall effect. All of the other three band
model is topologically equivalent to spin-1 model unless the band
gap is closed. As a realistic example, we introduce the OMN model on
Kagome lattice which has unit Chern number including the meron
structure in momentum space. In Sec. III, by introducing the nematic
operator, we construct the different type of three band model with
time-reversal symmetry which involves the (pseudo-)spin Hall effect.
For an specific model based on Kagome lattice, we perform the linear
response calculation of the (pseudo-)spin Hall conductivity. Then we
give an physical interpretation for our (pseudo-)spin Hall effect.
In Sec. IV we conclude with brief summary and discussions.

\section{Spin-1 model}

In general, arbitrary three-band Hamiltonian can be written as a
linear combination of eight Gell-Mann matrices $\lambda^a$
($a=1,\cdots,8$), ${\cal H}_{\v k} =\varepsilon_{\v k}\mathbb{I} +
\sum_{a=1}^8 d^a_{\v k} \cdot \lambda^a$, characterized by
eight-component real field $d^a_{\v k}$. Due to its complicate
structure, the exact analytical solution of general three band model
is difficult to obtain. By the way, as is well-known, Gell-Mann
matrices are the generators of the SU(3) Lie algebra which possess
as subgroups several sets of three matrices forming SU(2), or SO(3)
sub-algebra. In particular the subset that generates the SO(3)
rotation are the matrices of spin-1 operators. We will show that the
subset of 3$\times$3 Hamiltonians spanned by the SO(3) generators
can be readily analyzed in analogous fashion as their 2$\times$2
counterparts. We specialize to three-band Hamiltonians of the type
\begin{align}\label{Ham2}
{\cal H}_{\v k} =\v d_{\v k} \cdot \textbf{S},
\end{align}
where one possible choice of the spin-1 matrix $\textbf{S}$ would be
$(S^\alpha )_{\beta\gamma} = -i \varepsilon_{\alpha\beta\gamma}$, or $\v S = (\lambda^7, - \lambda^5, \lambda^2)$:
\ba &&S^x = \lambda^7 = \bpm 0 & 0 & 0 \\ 0 & 0 & -i \\ 0 & i & 0 \epm,~
S^y = -\lambda^5 =  \bpm 0 & 0 & i \\ 0 & 0 & 0 \\ -i & 0 & 0\epm,\nn
&&S^z = \lambda^2 = \bpm 0 & -i & 0 \\ i & 0 & 0 \\ 0 & 0 & 0\epm. \ea
All other choices of SO(3) subgroup elements ought to be related to this by a suitable
unitary transformation. As in the two-band Hamiltonian, we assume
$|\v d_{\v k}|$ remains nonzero throughout the BZ. The eigenstates
of energies $d_{\v k}$ and $0$ can be worked out readily in terms of
the unit vector $\hat{d}_{\v k} = (\hat{d}_x, \hat{d}_y, \hat{d}_z
)$~\cite{varma,comment1},
\ba | \psi^+_{\v k} \rangle\!=\!\frac{1}{\sqrt{2[ 1\!-\!( \hat{d}_z
)^2]}} \left(\begin{array}{c}
 {\hat d}_x {\hat d}_z + i {\hat d}_y  \\
 {\hat d}_y {\hat d}_z - i {\hat d}_x \\
 {\hat d}_z^2-1, \end{array}\right), ~
|\psi^0_{\v k} \rangle\!=\!\left(\begin{array}{c} {\hat d}_x   \\
{\hat d}_y  \\  {\hat d}_z \end{array}\right) \!.\label{eq:psi+psi0}
\ea
The state of energy $-d_{\v k}$ is the complex conjugate of
$|\psi^+_{\v k}\rangle$: $|\psi^-_{\v k} \rangle=(|\psi^+_{\v k}
\rangle)^\ast$.

With the aid of explicit wave functions one can evaluate the gauge
flux associated with each band. The central band, having real-valued
wave functions, has the zero flux, while the upper band has the flux
density\cite{varma}

\ba \rho_{\v k} = \left( {\partial a_y \over
\partial k_x } - {\partial a_x \over
\partial k_y } \right) = \hat{d}_{\v k} \cdot \left( {\partial
\hat{d}_{\v k} \over
\partial k_x} \times {\partial \hat{d}_{\v k} \over \partial k_y }
\right). \label{eq:spin1-flux}\ea
The lower band has the opposite sign of the flux density. Comparing Eq.
(\ref{eq:spin1-flux}) to Eq. \eqref{eq:Skyrmion-density} one notes a difference of factor
2, originating from spin-1 being twice the size of spin-1/2~\cite{varma}. A full Skyrmion for
$\hat{d}_{\v k}$ in the BZ will thus imply an even Chern number for the spin-1 bands.

\begin{figure}[ht]
\includegraphics[width=85mm]{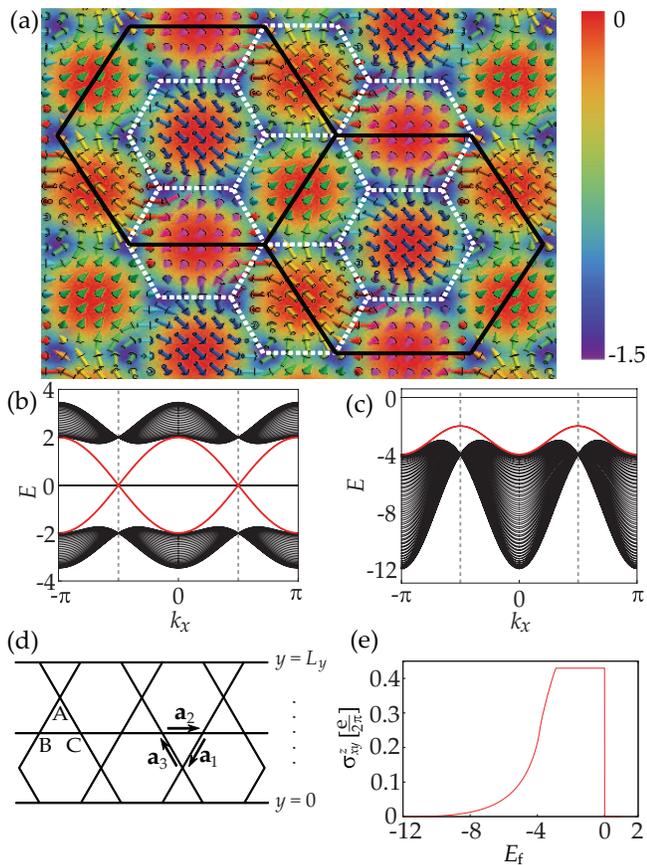}
\caption{(color online) (a) Skyrmion density (background color) for the parent
Hamiltonian of the OMN model. Color bar represents the Skyrmion density. The $\v d_{\v
k}$-vector (arrows) is periodic over four BZs as indicated by the black hexagon. White
hexagon is the BZ. (b) Energy dispersion for open boundary condition of the parent
Hamiltonian $\v d_{\v k}\cdot \v S$. The one-dimensional BZ is indicated by two dashed
vertical bars at $k_x =\pm \pi/2$. Edge modes are shown in red, bulk modes in black. (c)
Energy dispersion of the folded OMN Hamiltonian $-(\v d_{\v k}\cdot \v S)^2$ under the
open boundary condition. (d) Open geometry used in the calculation of (b) and (c).
Hopping occurs along the solid bonds only. Three unit vectors $\v a_1, \v a_2, \v a_3$
are shown. Three sublattices sites are indicated as A, B, C. (e) Spin Hall conductance
(in units of $e/2\pi$) for the folded OMN model as a function of the Fermi
energy.}\label{fig:OMN}
\end{figure}

We ask now if certain topological three-band models considered in
earlier literature can be framed in the form of the ``parent
Hamiltonian" $\v d_{\v k}\cdot \v S$ plus some perturbation. As long
as the parent model is connected smoothly to the full one without
the gap closing the topological property will be completely captured
by the relation, Eq. (\ref{eq:spin1-flux}). The Kagome lattice model
of OMN~\cite{OMN} is

\begin{align}\label{DH1}
{\cal H}^{(\mathrm{OMN})}_{\v k} = \v d_{\v k}\cdot \v S + \v d'_{\v
k} \cdot \v S' ,
\end{align}
where $\v S = (\lambda^7, -\lambda^5, \lambda^2 )$, $\v S' = (\lambda^6, \lambda^4,
\lambda^1 )$, while $\v d_\v k\!=\!2\sin [\phi/3] (\cos({\v k}\cdot {\v a_2}),\cos({\v k}\cdot {\v a_3}),\cos({\v k}\cdot {\v a_1}))$ and
$\v d'_\v k\!=\!2\cos[\phi/3](\cos({\v k}\cdot {\v a_2}),\cos({\v k}\cdot {\v a_3}),\cos({\v k}\cdot {\v a_1}))$
for the three unit vectors
$\v a_i$ $(i=1,2,3)$ of the Kagome lattice shown in Fig. \ref{fig:OMN}(d). The flux
$\phi$ penetrating the triangle unit of the Kagome lattice is compensated for by $-2\phi$
flux through the hexagon in the OMN model. We have verified that $\v d'_{\v k} \cdot \v
S'$ does not close the energy gap of the parent Hamiltonian $\v d_{\v k}\cdot \v S$,
provided the flux $\phi$ satisfies the condition ${\rm sgn}\left(\sin [\phi/3] \right) =
-{\rm sgn} (\sin \phi )$. The Skyrmion number associated with the $\v d_{\v k}$-vector of
the parent OMN Hamiltonian is readily obtained, $N_s =-(1/2) {\rm sgn}\left(\sin [\phi
/3] \right)$, giving rise to the Chern number for the upper band $C^+ = -{\rm
sgn}\left(\sin [\phi/3]\right)$ according to Eq. (\ref{eq:spin1-flux}). The Chern number
obtained for the same band in Ref. \onlinecite{OMN} is $C^+ = {\rm sgn}(\sin \phi)$,
equal to our result provided ${\rm sgn}\left(\sin [\phi/3] \right) = -{\rm sgn} (\sin
\phi )$. This is precisely the same condition required for the topological equivalence of
the original OMN model to its parent form. Written in real space lattice the parent
Hamiltonian gives the hopping amplitude proportional to $+i$ for every nearest neighbor
bond traversed in the counter-clockwise direction.

An astute reader may wonder how come the Skyrmion number in the
parent OMN Hamiltonian is only half of an integer. In fact any
Hamiltonian matrix $\v d_{\v k} \cdot \v S$ that shares the
periodicity of the first BZ, i.e. $\v d_{\v k + \v G} = \v d_{\v k}$
for reciprocal lattice vector $\v G$, will only allow the integer
Skyrmion numbers due to the homotopy $\pi_2 (S^2) = \mathbb{Z}$ and,
by virtue of Eq. (\ref{eq:spin1-flux}), only even Chern integers. On
the other hand, explicit calculation of the Berry phase flux yields
the Chern numbers of $\pm 1$~\cite{OMN}. So why the apparent
contradiction? A closer examination shows that $\v d_{\v k}$ for the
OMN model is not periodic under $\v k \rightarrow \v k + \v G$ but
rather under $\v k \rightarrow \v k + 2\v G$ as shown graphically in
Fig. \ref{fig:OMN}(a). The same relation between Chern number and
winding number for three orbital model with the spin-1
representation is studied in Ref. ~\onlinecite{varma}. As a result,
${\v d}_{\v k + \v G} \cdot \v S$ is equivalent to ${\v d}_{\v
k}\cdot \v S$ only up to some constant, $\v k$-independent unitary
rotation: $[ {\v d}_{\v k + \v G} \cdot \v S ] = U^\dag [ {\v d}_{\v
k} \cdot \v S ] U$. By evading the periodic condition the
Hamiltonian becomes exempt from the usual homotopy consideration as
well, making the half Skyrmion configuration possible. In fact with
a different choice of the basis the OMN matrix can be written in
such a way that ${\cal H}_{\v k + \v G} = {\cal H}_{\v k}$ holds. In
that case, however, the model can no longer be reduced to the form
$\v d_{\v k} \cdot \v S$ and the Skyrmion number interpretation of
the Chern number fails to apply~\cite{lee-comment}.

Topological nature of the spin-1 Hamiltonian ${\cal H}_{\v k} = \v
d_{\v k} \cdot \v S$ can be phrased in the general classification
scheme of Ref.~\onlinecite{tenfold-way}. Under complex conjugation
one has $\v S^* = -\v S$ and ${\cal H}_{\v k}^* = (\v d_{\v k} \cdot
\v S)^* = -{\cal H}_{\v k}$. The Hamiltonian possesses time-reversal
symmetry (TRS) and/or particle-hole symmetry (PHS) if some unitary
transformation could relate ${\cal H}_{\v k}^*$ to $+{\cal H}_{-\v
k}$ (TRS) or $-{\cal H}_{-\v k}$ (PHS), respectively. When the $\v
d_{\v k}$-vector is even, $\v d_{\v k} = \v d_{-\v k}$, we further
have ${\cal H}_{-\v k} = {\cal H}_{\v k}$, hence the TRS/PHS amounts
to the existence of unitary transformation $U^\dag {\cal H}_{\v k} U
= \mp {\cal H}_{\v k}$, respectively. It follows that SLS is
trivially satisfied with $U=\mathbb{I}$, while TRS cannot be
achieved with any $U$. This places our Hamiltonian (\ref{Ham2}) in
the D class~\cite{tenfold-way}, where the allowed topological
numbers are the integers in two spatial dimensions, equal to the
Chern numbers we just calculated. If instead we had $\v d_{-\v k}
=-\v d_{\v k}$ the Hamiltonian would have TRS but not PHS, placing
it in class AI, without any topological numbers. Indeed one can
easily show that for the $\v d_{\v k}$-vector of odd symmetry the
Skyrmion number vanishes identically.

\section{Nematic model}

So far the discussion of the topological character of three-band
Hamiltonian is restricted to the ``spin" type given by Eq.
(\ref{Ham2}). The nomenclature is obviously derived from $\v S$
being a representation of $S=1$ spin. On the other hand, the
following observation prompts us to study another class of
three-band Hamiltonians that we denote the ``nematic" type. Note
that the anti-commutators of the three spin operators $(S^x, S^y,
S^z)$ generate the following five:

\ba N_1 &=& S^x S^y + S^y S^x = \lambda^1 = \bpm 0 & 1 & 0 \\ 1 & 0 & 0 \\ 0 & 0 & 0\epm, \nn
N_2 &=&  S^y S^z + S^z S^y = \lambda^6 = \bpm 0 & 0 & 0 \\ 0 & 0 & 1 \\ 0 & 1 & 0 \epm, \nn
N_3 &=&  S^z S^x + S^x S^z = - \lambda^4 = \bpm 0 & 0 & -1 \\ 0 & 0 & 0 \\ -1 & 0 & 0\epm, \nn
N_4&=&  [S^x ]^2 - [S^y ]^2  =-\lambda^3 = \bpm -1 & 0 & 0 \\ 0 & 1 & 0 \\ 0 & 0 & 0 \epm, \nn
N_5 &=&  {1\over \sqrt{3}} (2 [S^z ]^2 - [S^x ]^2 - [S^y ]^2 ) =
\lambda^8 = {1 \over \sqrt{3}}\bpm 1 & 0 & 0 \\ 0 & 1 & 0 \\ 0 & 0 & -2 \epm.\nn \ea
These are precisely the remaining five of the Gell-Mann matrices, or in the language of
spin liquids, the nematic operators. An arbitrary 3-band Hamiltonian is therefore a sum
of the spin part and the nematic part, ${\cal H}_{\v k} = \v d_{\v k} \cdot \v S + \v
D_{\v k} \cdot \v N$, where the five components of $\v N$ refer to the above five
operators and $\v D_{\v k}$ is a five-component function of $\v k$. The Hamiltonian
consisting solely of the nematic operators $\v D_{\v k} \cdot \v N$ are real, $[\v D_{\v
k} \cdot \v N]^* = \v D_{\v k} \cdot \v N$, and TRS/PHS conditions become $U^\dag [ \v
D_{\v k} \cdot \v N ] U = \pm \v D_{-\v k} \cdot \v N$. In this case even (odd) $\v D_{\v
k} = + (-) \v D_{-\v k}$ generates the AI (D) class Hamiltonian~\cite{tenfold-way}.

A particularly simple kind of nematic Hamiltonian arises by folding
the previous spin Hamiltonian

\ba {\cal H}_{\v k} \!=\! \v d_{\v k}\cdot \v S \rightarrow {\cal
H}^{(f)}_{\v k} \!=\! - [\v d_{\v k}\cdot \v S]^2 . \label{eq:OH}\ea
The folding ensures the degeneracy of Kramers' pairs $|\psi^\pm_{\v
k}\rangle$ with the identical energy $- (\v d_{\v k})^2$ wherein
non-Abelian gauge connection
arises~\cite{wilczek,MNZ1,MNZ2,orbitronics}. The same wave functions
worked out before the folding remain as eigenstates of the new
Hamiltonian. A similar three-band Hamiltonian, with $\v d_{\v k}
\propto \v k$, was proposed for the $p$-orbital bands of
Si~\cite{orbitronics}. An example of the folded four-band model
employing the $S=3/2$ spin operator $\v S$  is the Luttinger
Hamiltonian, extensively studied by Murakami, Nagaosa, and Zhang as
a model for dissipationless spin Hall current in
GaAs~\cite{MNZ1,MNZ2}. Even for spinless models such as ours, an
analogue of spin Hall current can be defined and its response
function computed. The special case of $\v d_{\v k} \propto \v k$
was analyzed in Ref. \cite{orbitronics}. Here we maintain the
framework as general as possible by keeping $\v d_{\v k}$ an
arbitrary non-zero vector over the BZ.

Each band has the associated helicity number, $\hat{d}_{\v k} \cdot
\v S = h$ ($h=+1, 0,-1$) and the band-dependent flux density,

\begin{align}\label{fij}
\displaystyle{F_{ij, \v k }^h ={h} \left(\hat d_{\v
k}\cdot\frac{\partial \hat d_{\v k} }{\partial
k^i}\times\frac{\partial \hat d_{\v k}}{\partial k^j}\right)}.
\end{align}
The sum of Chern numbers for the doubly degenerate band is obviously
zero. Regarding $\v S$ as the pseudo-spin matrix, the projected
pseudo-spin density may be defined as~\cite{MNZ1,MNZ2,orbitronics}

\ba \label{consd} {\tilde\rho}^\alpha_{\v q} = \sum_{\v p}
\psi^\dag_{\v p + \frac{1}{2}\v q} {\tilde S}^\alpha_{\v p} \psi_{\v
p-\frac{1}{2}\v q}, ~ {\tilde S}^\alpha_{\v p} = \sum_{a=0,1}
P^{(a)}_{\v p} S^\alpha P^{(a)}_{\v p}, \ea
with $P^{(0)}_{\v p} =1-(\hat d_{\v p} \cdot \v S)^2$ and $P^{(1)}_{\v p} =(\hat d_{\v p}
\cdot \v S)^2$ denoting the projection onto the eigenstates with $E_0=0$ and $E_1=-(\v
d_{\v p} )^2$, respectively, and $\psi_{\v p}$ is the three-component spinor consisting
of A, B, C sublattice site operators. Together with the projected spin current operator

\begin{align}\label{prospo}
\v {\tilde J}^\alpha_{\v q} =\sum_{\v p} \psi^\dag_{\v p
+\frac{1}{2}\v q }\, \left\{ \tilde S^\alpha, \frac{\partial {\cal
H}^{(f)}_{\v p}}{\partial \v p} \right\}\, \psi_{\v p-\frac{1}{2}\v
q},
\end{align}
they obey the continuity equation $\dot{\tilde \rho}^\alpha_{\v q} =
-i\v q \cdot \v {\tilde J}^\alpha_{\v q}+ O(\v q^2)$. The d.c.
pseudo-spin Hall conductivity follows as
($V$=volume)~\cite{MNZ2,orbitronics}

\begin{align}\label{DCres}
\sigma_{ij}^\alpha = -\frac{1}{V}\sum_{\v k} d^\alpha_{\v k} F_{ij,
\v k}^{+1} \, \left[ n^{(1)}_{\v k} -n^{(0)}_{\v k}\right],
\end{align}
where $n^{(a)}$ is the Fermi function of each band $a=0,1$. Here $j$
stands for the direction of the applied electric field, $i$ the
spatial direction of spin current, and $\alpha$ is the spin
orientation. This formula is general and applicable to any
three-band Hamiltonians $H^{(f)}_{\v k}$ of the folded form, Eq.
(\ref{eq:OH}). Numerical evaluation of the pseudo-spin Hall
conductance $\sigma_{xy}^x $ for the parent OMN model is shown in
Fig. \ref{fig:OMN}(e). Due to symmetry other spin orientations give
the same Hall conductance: $\sigma^x_{xy} = \sigma_{xy}^y =
\sigma_{xy}^z$.

\begin{figure}[ht]
\includegraphics[width=80mm]{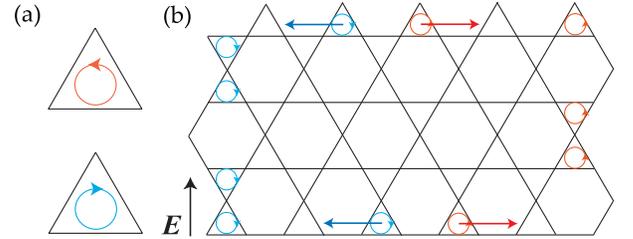}
\caption{(color online) (a) Two opposite current loops around a ABC triangle.
(b) Pseudo-spin Hall phenomena. Two opposite loops move in the opposite
directions. The electric field is applied along vertical axis.}
\label{fig:PSHE}
\end{figure}

Intuitive understanding of the pseudo-spin Hall conductance follows
from the fact that spin operator we use also serves as the
sublattice current operator. For instance, $S^x$ expressed in the
sublattice basis $|\mathrm{A}\rangle, |\mathrm{B}\rangle,
|\mathrm{C}\rangle$ becomes $S^x = -i |\mathrm{B}\rangle \langle
\mathrm{C}| +i |\mathrm{C} \rangle \langle \mathrm{B}|$, equal to
orbital current operator $J^\mathrm{o}_\mathrm{BC}$ for that bond.
As a whole we may identify $\v S \equiv (J^\mathrm{o}_\mathrm{BC},
J^\mathrm{o}_\mathrm{CA}, J^\mathrm{o}_\mathrm{AB})$ in obvious
notation. The eigenstate $|\psi^+_{\v k}\rangle$ has the average
$\langle \psi^+_{\v k} | \v S | \psi^+_{\v k} \rangle = \hat{d}_{\v
k}$, and with our new interpretation it implies that the current
loop around a ABC triangle shown in Fig. \ref{fig:PSHE}(a) is
$\langle\psi^+_{\v k} |(
J^\mathrm{o}_\mathrm{AB}+J^\mathrm{o}_\mathrm{BC}+J^\mathrm{o}_\mathrm{CA}
)| \psi^+_{\v k} \rangle = \hat{d}_x + \hat{d}_y + \hat{d}_z$ and
minus this value for $|\psi^-_{\v k}\rangle$. These opposite current
loops are the analogues of opposite spin orientations in the genuine
spin Hall effect. Now these current loops of opposite signs move in
the opposite directions due to the opposite signs of the gauge flux
they experience, conserving the time-reversal invariance, and result
in the pseudo-spin Hall phenomena (Fig. \ref{fig:PSHE}(b)).

Imposing boundaries in the $y$-direction at $y=0$ and at $y=L_y$ as
shown in Fig. \ref{fig:OMN}(d)  introduces some edge modes. The
topologically protected edge modes~\cite{hatsugai} of the parent OMN
model connecting the topological bands (top and bottom) with the
non-topological band (center) as shown in Fig. \ref{fig:OMN}(c) obey
the dispersion~\cite{leHur} $E_{k_x} =\pm 2 \cos k_x $ when the
hopping magnitude is chosen to one. The folded Hamiltonian $-(\v
d_{\v k} \cdot \v S)^2$ also support edge modes, albeit unprotected
in the topological sense, with the dispersion $E_{k_x} =-(3+\cos
2k_x)$ as shown in Fig. \ref{fig:OMN}(d). The explicit derivation of
the edge dispersion is presented in the Appendix. For the OMN model
the sign of the velocity $dE_{\v k_x}/dk_x$ determines the spatial
location of the edge as either $y\simeq 0$ or $y\simeq L_y$. In the
folded case both edge modes obey the same energy dispersion. The
absence of protected edge mode connecting the degenerate bands to
the flat band is due to the net topological number for the former
bands being zero~\cite{hatsugai,ON}.

\section{Conclusion and Discussion}
In this paper we have taken up the study of the topological aspects
of three-band Hamiltonian, applicable to spinless fermions and
bosons with three-fold orbital or sublattice degree of freedom. As
with the topological two-band models, the Hall conductances of the
three-band Hamiltonian is shown to be governed by the same (Chern
number) $\leftrightarrow$ (Skyrmion number) correspondence with a
critical difference of factor two allowing for the existence of
meron structure in momentum space for three-band Hamiltonians. For
the OMN model we succeeded in obtaining the topological number of the
model by using the parent model of Eq. (\ref{Ham2}). However, we
didn't mention about how to obtain the parent model from the general
three band model. In this point, we need further studies. A second
class of three-band Hamiltonians obtained by the folding procedure
is studied. The double degeneracy ensured by folding naturally leads
to non-Abelian gauge structure and spin Hall phenomena. Both of
three-band models we study, before and after the folding, are
subject to the symmetry classification scheme of Ref.
\onlinecite{tenfold-way}.

\acknowledgments J. H. H. is supported by NRF grant (No.
2010-0008529, 2011-0015631). We acknowledge informative discussions
with H. Katsura, Dung-Hai Lee, E. G. Moon, and N. Nagaosa.

\appendix*

\setcounter{equation}{0}

\section{Edge modes in the folded model}
In the OMN model, the chiral edge state is obtained in Ref.
\onlinecite{leHur}. Here we calculate the edge mode solutions of the
folded OMN model,
\begin{align}
{\cal
H}^{(f)}_{\v k} \!=\! - [\v d_{\v k}\cdot \v S]^2,
\end{align}
where
\begin{align}
\v d_\v k\!=\!2\sin [\phi/3] (\cos({\v k}\cdot {\v a_2}),\cos({\v k}\cdot {\v a_3}),\cos({\v k}\cdot {\v a_1})).
\end{align}
The real space expression of the Hamiltonian is obtained by writing down its
Fourier transformation,
%
\ba && H^{(f)} = -\sum_{\v r} \Bigl( a^{\dag}_{\v r} [a_{\v r+ 2 \v
a_3} \!+\! a_{\v r+2 \v a_1} ] \nn
&&~+ b^{\dag}_{\v r} [b_{\v r+ 2 \v a_1} \!+\! b_{\v r+2 \v a_2} ]
\!+\! c^{\dag}_{\v r} [c_{\v r+ 2 \v a_2} \!+\! c_{\v r+2 \v a_3} ]
\nn
&&~- a^{\dag}_{\v r} [b_{\v r + \v a_3- \v a_2} \!+\! b_{\v r + \v
a_1} \!+\! c_{\v r + \v a_3} \!+\! c_{\v r + \v a_1-\v a_2}]\nn
&&~- b^{\dag}_{\v r} [ c_{\v r + \v a_2} \!+\! c_{\v r + \v a_3-\v
a_1} \!+\! a_{\v r + \v a_3- \v a_2} \!+\! a_{\v r + \v a_1} ]\nn
&&~- c^{\dag}_{\v r} [a_{\v r + \v a_1- \v a_2} \!+\! a_{\v r + \v
a_3} \!+\! b_{\v r + \v a_2} \!+\! b_{\v r + \v a_3-\v a_1}]
\Bigr)\!+\! h.c. \nn \label{eq:folded-OMN} \ea
In order to obtain the
edge states, we should choose a particular boundary condition.
The boundary condition is depicted in Fig. \ref{fig:OMN} (d).
Since the system is no longer periodic along the $y$ direction,
$k_y$ is not good quantum number. Let us consider momentum representation in $x$ direction
\begin{align}
a(\v r)=a_{n_1,n_2}=\frac{1}{\sqrt{N_x}} \sum_{k} e^{-i k (2n_1+n_2)} a_{n_2}(k).
\end{align}
Here, we used $\v r=(2n_1+n_2, \sqrt{3} n_2)$.
Inserting the one particle state
\begin{widetext}
\begin{align}
|\Psi(k)\rangle= \sum_{j=1} \left( \psi^a_j (k) a^\dag_j(k) +\psi^b_{j-\frac12} (k) b^\dag_{j-\frac12}(k) +\psi^c_{j-\frac12} (k) c^\dag_{j-\frac12}(k)\right)|0\rangle
\end{align}
into the Schrodinger equation $H|\Psi\rangle= E|\Psi\rangle$, we have the one-dimensional chain equations of
$\psi_j$~\cite{hatsugai}
\begin{align}
&E\psi^a_j= 2\cos k \left(\psi^a_{j+1}+ \psi^a_{j-1} - e^{\frac{i}{2}k} \psi^b_{j+\frac12}- e^{-\frac{i}{2}k} \psi^b_{j-\frac12}
- e^{-\frac{i}{2}k} \psi^c_{j+\frac12}- e^{\frac{i}{2}k} \psi^c_{j-\frac12}\right),\label{ceq2-1}\\
&E\psi^b_{j-\frac12}= 2\cos(2k) \psi^b_{j-\frac12}+ e^{ik}\psi^b_{j-\frac32} + e^{-ik} \psi^b_{j+\frac12}\nn
&\hspace{15mm}- 2\,e^{-\frac{i}{2}k} \cos k \,\psi^a_{j-1} - 2\, e^{\frac{i}{2}k} \cos k\, \psi^a_{j}- 2\cos k \, \psi^c_{j-\frac12}-\psi^c_{j+\frac12}-\psi^c_{j-\frac32},\label{ceq2-2}\\
&E\psi^c_{j-\frac12}= 2\cos(2k) \psi^c_{j-\frac12}+ e^{-ik}\psi^c_{j-\frac32} + e^{ik} \psi^c_{j+\frac12}\nn
&\hspace{15mm}- 2\,e^{\frac{i}{2}k} \cos k \,\psi^a_{j-1} - 2\, e^{-\frac{i}{2}k} \cos k\, \psi^a_{j}- 2\cos k \, \psi^b_{j-\frac12}-\psi^b_{j+\frac12}-\psi^b_{j-\frac32},\label{ceq2-3}
\end{align}
\end{widetext}
with the boundary conditions
\begin{align}
&\psi^a_0=\psi^b_{-\frac12}=\psi^c_{-\frac12}=0,\qquad \psi^b_{\frac12}=1,\nn
&\psi^a_{\rm{max+1}}=\psi^b_{\rm{max+\frac12}}=\psi^c_{\rm{max+\frac12}}=0.
\end{align}
For exponentially localized solutions on the boundaries we take the ansatz~\cite{creutz}
\begin{align}\label{aned}
\psi^a_j= \eta_a^{j-1} \phi^a, \qquad \psi^b_{j-\frac12}=\eta_b^{j-1} \phi^b, \qquad \psi^c_{j-\frac12}=\eta_c^{j-1} \phi^c,
\end{align}
with the condition $|\psi^b|=|\psi^c|$. Thus we can write as
\begin{align}
\phi^b=1,\qquad \phi^c=e^{i\chi},\qquad \eta_{b,c}=\eta e^{i \theta_{b,c}},
\end{align}
where $\eta$ is real function of momentum $k$.
Putting this all together we obtain the edge solutions as
\begin{align}
\psi^a_j=0, \qquad \psi^b_{j-\frac{1}{2}}=\eta^{j-1},\qquad \psi^c_{j-\frac{1}{2}}=\eta^{j-1} e^{\pm ik},
\end{align}
with
\begin{align}
\eta=-(\cos k)^{\pm 1}, \qquad E_{k} =-(3+\cos 2k).
\end{align}
The degenerate edge spectrum are depicted in Fig.~\ref{fig:OMN} (c).
For $\eta^2 >1$($\eta^2 <1$), the edge state is localized at upper(lower) edge.
\begin{widetext}
\begin{figure}[ht]
\includegraphics[width=80mm]{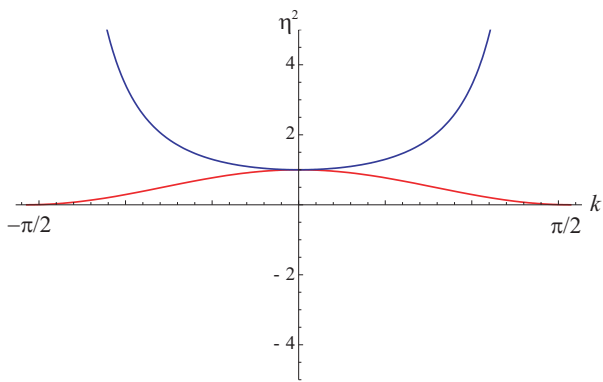}
\caption{(color online) Plot of $\eta^2$. For $\eta^2 >1$($\eta^2 <1$),
the edge state is localized at upper(lower) edge (blue(red) line).}
\label{fig:lambda}
\end{figure}
\end{widetext}


\begin{thebibliography}{99}
\bibitem{comment1} The wave function for $|\psi^+_{\v k} \rangle$ contains a singularity
at $d_z = \pm 1$. Another wave function can be easily constructed for $|\psi^+_{\v k}
\rangle$ where the singularity appears elsewhere on the unit sphere\cite{varma}.
\end{thebibliography}
\end{document}